\documentclass[conference]{IEEEtran}
\ifCLASSINFOpdf
 \usepackage[pdftex]{graphicx}
 \usepackage{amsfonts}
 \usepackage{amssymb}
 \usepackage{color}
\usepackage{eurosym}
\DeclareGraphicsExtensions{.pdf}
\else
\fi
\begin{document}
\title{Economic networks in and out of equilibrium}



\author{\IEEEauthorblockN{Tiziano Squartini, Diego Garlaschelli}
\IEEEauthorblockA{Instituut-Lorentz for Theoretical Physics\\
Leiden Institute of Physics, University of Leiden\\
Niels Bohrweg 2, 2333 CA Leiden (The Netherlands)\\
Email: squartini@lorentz.leidenuniv.nl, garlaschelli@lorentz.leidenuniv.nl}
}

%\author{\IEEEauthorblockN{Francesco Picciolo, Franco Ruzzenenti, Riccardo Basosi}
%\IEEEauthorblockA{Department of Chemistry\\
%University of di Siena\\
%Siena, Via Aldo Moro 1, 53100 Siena (Italy)\\
%Email: picciolo@unisi.it, ruzzenenti@unisi.it, basosi@unisi.it}
%\and
%\IEEEauthorblockN{Tiziano Squartini, Diego Garlaschelli}
%\IEEEauthorblockA{Instituut-Lorentz for Theoretical Physics\\
%Leiden Institute of Physics, University of Leiden\\
%Niels Bohrweg 2, 2333 CA Leiden (The Netherlands)\\
%Email: squartini@lorentz.leidenuniv.nl}
%}


\maketitle

\begin{abstract}
Economic and financial networks play a crucial role in various important processes, including economic integration, globalization, and financial crises.
Of particular interest is understanding whether the temporal evolution of a real economic network is in a (quasi-)stationary equilibrium, i.e. characterized by smooth structural changes rather than abrupt transitions.
Smooth changes in quasi-equilibrium networks can be generally controlled for, and largely predicted, %via an appropriate rescaling of structural quantities, 
while this is generally not possible for abrupt transitions in non-stationary networks.
Here we study whether real economic networks are in or out of equilibrium by checking their consistency with quasi-equilibrium maximum-entropy ensembles of graphs.
As illustrative examples, we consider the International Trade Network (ITN) and the Dutch Interbank Network (DIN).
We show that, despite the globalization process, the ITN is an almost perfect example of quasi-equilibrium network, while the DIN is clearly an out-of-equilibrium network undergoing major structural changes and displaying non-stationary dynamics. 
Among the out-of-equilibrium properties of the DIN, we find striking early-warning signals of the interbank crisis of 2008.
\end{abstract}



% \fi
\IEEEpeerreviewmaketitle

\section{Introduction}

Economic and financial systems are strongly interconnected, with several units being linked to each other via different possible types of interaction. Important examples include trade networks, where pairs of economic agents exchange goods or services in return of money, and lending networks, where financial institutions lend and borrow money from each other giving rise to another kind of directed relationship.

In general, understanding the (typically intricate and irregular) structure of economic networks is crucial in order to study economic dynamics, especially under stress conditions. For instance, trade networks (especially at the worldwide level) are an important ingredient of economic integration and globalization \cite{IEEEhowto:giorgio,IEEEhowto:giorgio2}, while lending networks play a central role for financial (in)stability at an international level \cite{IEEEhowto:preferentiallending}. 
In particular, the recent global financial and interbank crisis has witnessed how the collapse of the system was due to an increased interconnectedness of the interbank network \cite{IEEEhowto:may1,IEEEhowto:subprime}. While individual banks felt safe minimizing their individual risk by diversifying their portfolios, the simultaneous diversification of the portfolios of all banks resulted in an unexpected and uncontrolled level of mutual dependency, thereby amplifying the effects of individual defaults \cite{IEEEhowto:may1,IEEEhowto:subprime}.

%The above considerations imply that it is very important to study economic networks from a dynamical point of view. 
A particularly interesting question is whether the temporal evolution of a given economic network is (quasi-)stationary, i.e. if the system undergoes smooth structural changes, as opposed to abrupt transitions or other signatures of non-stationary dynamics.
Smooth changes in quasi-stationary networks can be generally controlled for, %via an appropriate rescaling of structural quantities, with an obvious benefit in terms of a resulting simplified and largely predictable description of the system's evolution, 
while this is generally not possible for abrupt transitions in non-stationary networks.

In this paper, we address the (non-)stationarity of real economic networks by studying whether they are found to be typical members of an evolving \emph{quasi-equilibrium ensemble of graphs} with given properties \cite{IEEEhowto:shannon,IEEEhowto:jaynes,IEEEhowto:HL,IEEEhowto:WF,IEEEhowto:newman_expo,IEEEhowto:mylikelihood,IEEEhowto:mymethod}. Roughly speaking, we identify a set of graph properties that we expect to change in time as the natural result of the internal evolution of the economic units (the nodes of the network), and we check whether the evolution of the entire network can be simply explained in terms of the changes in the selected properties. Such properties are treated as \emph{constraints} \cite{IEEEhowto:jaynes,IEEEhowto:HL,IEEEhowto:WF,IEEEhowto:newman_expo,IEEEhowto:mylikelihood,IEEEhowto:mymethod}, since we assume that they are in some sense the independent variables that can undergo an autonomous evolution, while the other properties of the network can vary only as long as they do not violate the independent variables \cite{IEEEhowto:newman_expo,IEEEhowto:mylikelihood,IEEEhowto:mymethod,IEEEhowto:mywtw}. 
If, over time, the network properties are systematically found to be in agreement with what expected merely from the enforced constraints, we can then conclude that the network is (consistent with) a quasi-stationary one driven by the dynamics of the constraints. If the network slightly deviates from the equilibrium expectations, but the deviating patterns are the same at all times and of the same entity, then the network can still be considered consistent with a quasi-stationary one, but in this case it is not completely driven by the dynamics of the constraints. By contrast, if the network deviates significantly from the quasi-equilibrium expectation and if the deviating patterns are different at different times, then it must be considered a non-stationary one.

We will consider two case studies: the International Trade Network (ITN), defined as the global network of world countries connected by directed import/export relationships (of which we analyse 6 yearly snapshots spanning five decades, i.e. 1950, 1960, 1970, 1980, 1990 and 2000) \cite{IEEEhowto:gleditsch}, and the Dutch Interbank Network (DIN), defined as the network of Dutch banks connected by directed lending/borrowing relationships (of which we consider 44 quarterly snapshots spanning 11 years from 1998 to 2008) \cite{IEEEhowto:mybanks}. 
For simplicity, we will consider both networks in their purely binary representation, i.e. as graphs where links are either present or absent, regardless of their magnitude.
We will show that, during the time interval considered, the ITN is an almost perfect example of quasi-stationary economic network, with trade patterns being in systematic agreement with an equilibrium ensemble of graphs specified only by local properties \cite{IEEEhowto:pre1,IEEEhowto:mymotifs}. By contrast, the DIN is a clear example of non-stationary network undergoing major structural changes and displaying different dynamical regimes \cite{IEEEhowto:mybanks}. Among the non-stationary properties of this network, we find striking early-warning signals of the global crisis of 2008.

\section{Quasi-equilibrium graph ensembles}
In this section we introduce, in abstract terms, the formalism that we use in order to study the stationarity of real networks.  Being general, this formalism does not uniquely apply to economic networks, but to any evolving network.

Let us first consider a single (static) snapshot of a real network. 
Any network can be uniquely specified by its \emph{adjacency matrix} $\mathbf{A}$, with entries $a_{ij}=1$ if a link from node $i$ to node $j$ is there, and $a_{ij}=0$ otherwise.
Let us denote the real network by the particular matrix $\mathbf{A}^*$.
Given a set of topological properties that we choose as constraints (symbolically, we denote such set of constraints with the vector notation $\vec{C}$), it is possible to construct a collection, or \emph{statistical ensemble}, $\mathcal{G}$, of graphs such that the average value $\langle\vec{C}\rangle$ of the constraints $\vec{C}$ over the graph ensemble is equal to the value $\vec{C}^*$ observed on the real network $\mathbf{A}^*$ \cite{IEEEhowto:mylikelihood,IEEEhowto:mymethod}.
The most satisfactory way to construct this ensemble is that of assigning each graph $\mathbf{A}$ a probability $P(\mathbf{A})$ such that Shannon's entropy
\begin{equation}
S\equiv -\sum_\mathbf{A\in \mathcal{G}}P(\mathbf{A})\ln P(\mathbf{A})
\end{equation}

\noindent is maximized, under the constraint
\begin{equation}
\langle\vec{C}\rangle=\sum_\mathbf{A\in \mathcal{G}}P(\mathbf{A})\vec{C}(\mathbf{A})=\vec{C}^*
\label{exp}
\end{equation}
where $\vec{C}(\mathbf{A})$ denotes the value of the properties $\vec{C}$ measured on the particular graph $\mathbf{A}$.
%The maximization of Shannon's entropy ensures that the resulting ensemble is the least biased, given the information specified by the constraints. In other words, the graph ensemble is the maximally random one among those that display the selected properties. 
The solution of the maximization of Shannon's entropy is the exponential distribution \cite{IEEEhowto:shannon}
\begin{equation}
P(\mathbf{A}|\vec{\theta})=\frac{e^{-H(\mathbf{A},\:\vec{\theta})}}{Z(\vec{\theta})},
\end{equation}

\noindent where the so-called \emph{Hamiltonian} $H(\mathbf{A},\:\vec{\theta})\equiv\vec{\theta}\cdot\vec{C}(\mathbf{A})$ is a linear combination of the chosen constraints and the normalization constant, or \emph{partition function}, is given by $Z(\vec{\theta})\equiv\sum_{\mathbf{A}\in\mathcal{G}}e^{-H(\mathbf{A},\:\vec{\theta})}$ \cite{IEEEhowto:jaynes,IEEEhowto:HL,IEEEhowto:WF,IEEEhowto:newman_expo,IEEEhowto:mymethod}.
The (initially unknown) parameters $\vec{\theta}$ are Lagrange multipliers ensuring that the expected value of each constraint can be set equal to the observed value, as prescribed by eq.(\ref{exp}).
The particular value $\vec{\theta}^*$ that realizes eq.(\ref{exp}) can be shown to be also the value that maximizes the log-likelihood  $\ln\mathcal{L}(\vec{\theta})=\ln P(\mathbf{A}^*|\vec{\theta})$
\cite{IEEEhowto:mylikelihood,IEEEhowto:mymethod}. 
Once the unknown parameters have been found, it is possibile to evaluate the expected value of any other topological quantity of interest, $X$, as follows 

\begin{equation}
\langle X\rangle^*=\sum_{\mathbf{A}\in\mathcal{G}}X(\mathbf{A})P(\mathbf{A}|\vec{\theta}^*).
\label{eq:X}
\end{equation} 

We have recently proposed a completely analytical method that %implements the above prescriptions in the fastest possible time %Our method 
allows to compare in the fastest possible time any topological property of the real network with the corresponding expected value over the constructed ensemble \cite{IEEEhowto:mymethod}.
If, after this comparison, the real network is found to be a typical member of the ensemble, one can conclude that the knowledge of the quantities chosen as constraints is enough in order to reproduce the original network.
Otherwise the knowledge of additional properties is required.
%So the abstract construction of the artificial ensemble is a sophisticated tool to answer an actually very simple and practical question: can the structure of the entire real network be explained in terms of a few key quantities? 

We now show how it is possible to extend the above ideas to study whether a \emph{dynamically evolving} network is consistent with a quasi-equilibrium ensemble.
Given a temporal sequence $\{\mathbf{A}^*(t)\}$ of snapshots of a real network and a set of constraints $\vec{C}$, for each timestep $t$ there is a different observed value $\vec{C}^*(t)$. The procedure described above can then be repeated in order to generate, for each timestep $t$, a different maximum-entropy graph ensemble such that the ensemble average $\langle \vec{C}(t)\rangle$ equals $\vec{C}^*(t)$.
In a straightforward  manner, it is possible to check whether each snapshot of the real network is well reproduced by the corresponding ensemble.

Now, we should recall that we are interested in understanding whether the evolution of the real network is consistent with a quasi-equilibrium process where the changes in the network's structure are (entirely) driven by smooth changes in only a small set of its topological properties.
This can be easily done by taking $\vec{C}(t)$ to be precisely the temporal sequence of desired properties, and using the procedure described above to check whether the real network's evolution is consistent with that of the quasi-equilibrium ensemble generated by the dynamics of $\vec{C}^*(t)$.

%It is important to stress that, for our purposes, 
The ideal set of driving properties should in general include only \emph{local} quantities directly attributed to some node-specific feature.
The most important example %(for an undirected network where links have no orientation) 
is when the driving property is the \emph{degree sequence}, i.e. the number of links of each node in the network. 
If $k_i$ denotes the degree, or number of links, of node $i$, then the vector $\vec{k}$ denotes the degree sequence of the entire network. Specifying the degree sequence as the driving quantity then amounts to choose
$\vec{C}(t)\equiv \vec{k}(t)$.
Being a completely local property, the degree of a node is the quantity most directly influenced by some `intrinsic' %(non-topological)
 feature of that node%, such as (in the economic examples) its wealth, income, capitalization, etc
. For instance, it has been shown that the degree of countries in the ITN is strongly correlated with the Gross Domestic Product (GDP) \cite{IEEEhowto:mywtw}.
Therefore, when asking whether an economic network undergoes a quasi-equilibrium evolution, %driven by the dynamics of its local topological properties
we are effectively asking whether the network evolution is driven by the changes of intrinsic economic variables%, without the need to actually measure the latter and study their explicit impact on the topology
.

Since the economic networks that we consider in this paper are \emph{directed}, i.e. they contain links with a given orientation, we should briefly discuss what are the possible local properties, generalizing the concept of degree sequence.%, that can be defined in a directed network. This leads us the introduction of two different ensembles.

\subsection{Directed Configuration Model}
In a directed network $\mathbf{A}$, for each node $i$ one can separately define the number $k^{out}_i=\sum_{j(\ne i)}a_{ij}$ of out-going links, or \emph{out-degree}, and the number  $k^{in}_i=\sum_{j(\ne i)}a_{ji}$ of in-going links, or \emph{in-degree}.
The in- and out-degree are the simplest node-specific local properties.
In an economic context, they represent the number of in-coming and out-going partners (respectively) of an agent, institution, or country. 
As they often reflect some nontrivial node-specific dynamics, these numbers are typically extremely heterogeneous in real economic networks \cite{IEEEhowto:preferentiallending}. It is therefore essential to inspect the higher-order network properties only after the in- and out-degree of all nodes have been carefully controlled for,
%For an evolving economic network, repeating this procedure at every timestep allows one 
to check whether the observed changes in the higher-order structure can be traced back to the variations of the in- and out-degrees.
If this is the case, one can conclude that the observed evolution of the entire system can be reabsorbed in the local dynamics of the degrees%, and that the latter encapsulate the driving economic factors shaping the network
.

If the in- and out-degree of all nodes are both included as constraints in the vector $\vec{C}$, one obtains the so-called \emph{Directed Configuration Model} (DCM), one of the most frequently used ensembles in the complex networks literature \cite{IEEEhowto:newman_expo}. The DCM Hamiltonian is

\begin{equation}
H(\mathbf{A},\vec{\theta})=\sum_{i=1}^N(\alpha_i k_i^{out}+\beta_i k_i^{in})
\end{equation} 

\noindent and the resulting probability coefficient for the generic network, $\mathbf{A}$, simply factorizes as a product over pairs of nodes:
\begin{equation}
P(\mathbf{A}|\vec{\theta})=\prod_{i}\prod_{j(\neq i)}p_{ij}^{a_{ij}}(1-p_{ij})^{1-a_{ij}}
\end{equation} 

\noindent where $p_{ij}\equiv\frac{x_{i}y_{j}}{1+x_{i}y_{j}}$ with $x_i\equiv e^{-\alpha_i}$, $y_i\equiv e^{-\beta_i}$ \cite{IEEEhowto:mymethod}. 
Given a real network $\mathbf{A}^*$, the parameters $\{x_i\}$ and $\{y_i\}$ can be set to the values $\{x_i^*\}$ and $\{y_i^*\}$ that maximize the likelihood of $\mathbf{A}^*$, or equivalently that enforce eq.(\ref{exp}). The latter reads in this case

\begin{eqnarray}
\left\{ \begin{array}{l}
\langle k_i^{out}\rangle=\sum_{j(\neq i)}\frac{x_i^*y_j^*}{1+x_i^*y_j^*}={k_i^{out}}^*\:\forall\:i\\
\langle k_i^{in}\rangle=\sum_{j(\neq i)}\frac{x_j^*y_i^*}{1+x_j^*y_i^*}={k_i^{in}}^*\:\forall\:i.
       \end{array} \right.
\label{dcmsys}
\end{eqnarray}

Once the unknown variables are numerically determined, the expected value of any adjacency matrix entry simply becomes $\langle a_{ij}\rangle^*=p_{ij}^*=\frac{x_{i}^*y_{j}^*}{1+x_{i}^*y_{j}^*}$. 
The latter can be used to immediately calculate the expected value $\langle X\rangle^*$ of any topological quantity $X$ of interest %(without requiring the unfeasible summation over all graphs as in the formal definition (\ref{eq:X}) 
\cite{IEEEhowto:mymethod}. %This expected value can then be compare it with the corresponding observed one, to check whether the real network is a typical member of the ensemble. 

\subsection{Reciprocal Configuration Model}
A more stringent choice of local properties in directed networks is one that distinguishes between \emph{reciprocated} and \emph{non-reciprocated} links.
A link from node $i$ to node $j$ is said to be reciprocated if the opposite link from $j$ to $i$ is also present in the network. 
Otherwise it is said to be non-reciprocated.
For a given node $i$, we might separately count the number $k_{i}^{\rightarrow}$ of non-reciprocated out-going links, the number $k_{i}^{\leftarrow}$ of non-reciprocated in-coming links, and the number $k_{i}^{\leftrightarrow}$ of reciprocated (out-going and in-coming at the same time) links. 
Mathematically, these three different `degrees' are defined as $k_{i}^{\rightarrow}\equiv\sum_{j(\neq i)}a_{ij}^{\rightarrow}$,  $k_{i}^{\leftarrow}\equiv\sum_{j(\neq i)}a_{ij}^{\leftarrow}$ and $k_{i}^{\leftrightarrow}\equiv\sum_{j(\neq i)}a_{ij}^{\leftrightarrow}$ respectively, where $a_{ij}^{\rightarrow}\equiv a_{ij}(1-a_{ji})$, $a_{ij}^{\leftarrow}\equiv a_{ji}(1-a_{ij})$ and $a_{ij}^{\leftrightarrow}\equiv\sum_{j(\neq i)}a_{ij}a_{ji}$. 

The graph ensemble where each of the above three quantities is specified for every node is known as the \emph{Reciprocal Configuration Model} (RCM)  \cite{IEEEhowto:mymethod,IEEEhowto:mygrandcanonical,IEEEhowto:myreciprocity}. 
Note that, once the three generalized degrees $k_{i}^{\rightarrow}$, $k_{i}^{\leftarrow}$ and $k_{i}^{\leftrightarrow}$ are specified, the `simpler' out- and in-degrees $k_{i}^{out}$ and $k_{i}^{in}$ are automatically specified as well, but the opposite is not true.
In an economic setting, %the reason for introducing this more stringent ensemble is the fact that 
the reciprocity of economic interactions reflects important properties, such as trust or preference.
Separately controlling for reciprocated and non-reciprocated interations means additionally controlling for the heterogeneity of these properties of nodes, besides the size heterogeneity already reflected in the in- and out-degrees. 

The Hamiltonian defining the RCM is the following:
\begin{eqnarray}
H(\mathbf{A},\vec{\theta})=\sum_{i=1}^N(\alpha_i k_i^\rightarrow+\beta_i k_i^\leftarrow+\gamma_i k_i^\leftrightarrow).
\end{eqnarray}

\noindent Even if the constraints are now non-linear combinations of the adjacency matrix entries,
the probability still factorizes as a product of dyadic probabilities, making the model analitically solvable \cite{IEEEhowto:mymethod,IEEEhowto:mymotifs,IEEEhowto:mygrandcanonical,IEEEhowto:myreciprocity}.
The maximization of the likelihood function leads to the following system of equations:

\begin{eqnarray}
\left\{ \begin{array}{l}
\langle k_i^{\rightarrow}\rangle=\sum_{j(\neq i)}\frac{x_i^*y_j^*}{1+x_i^*y_j^*+x_j^*y_i^*+z_i^*z_j^*}={k_i^{\rightarrow}}^*\quad\forall\:i\\
\langle k_i^{\leftarrow}\rangle=\sum_{j(\neq i)}\frac{x_j^*y_i^*}{1+x_i^*y_j^*+x_j^*y_i^*+z_i^*z_j^*}={k_i^{\leftarrow}}^*\quad\forall\:i\\
\langle k_i^{\leftrightarrow}\rangle=\sum_{j(\neq i)}\frac{z_i^*z_j^*}{1+x_i^*y_j^*+x_j^*y_i^*+z_i^*z_j^*}={k_i^{\leftrightarrow}}^*\quad\forall\:i
       \end{array} \right.
\label{rcmsys}
\end{eqnarray}
where $x_i\equiv e^{-\alpha_i}$, $y_i\equiv e^{-\beta_i}$, $z_i\equiv e^{-\gamma_i}$.

The addenda in the three equations above correspond to three different probability coefficients, that we denote as $(p_{ij}^{\rightarrow})^*$, $(p_{ij}^{\leftarrow})^*$ and $(p_{ij}^{\leftrightarrow})^*$ respectively.
These coefficients separately specify the probability of having, from node $i$ to node $j$, a non-reciprocated out-going link, a non-reciprocated in-coming link, and two reciprocated links respectively.
%As in the DCM, these probabilities can be used to calculate the expected value of the topological properties of interest.

\section{Triadic motifs: $z$-scores and significance profiles}
Since they assume that the network arises as a %(conceptually) 
simple combination of purely local properties, the DCM and RCM are typically treated as \emph{null models}.
Null models are simple models that one expects to fail in reproducing the data, and useful precisely because they can highlight interesting patterns in the real system in terms of deviations from the null hypothesis.
In the context we are considering here, the systematic accordance with a null model throughout the time period considered would indicate a quasi-equilibrium network evolution driven by the constraints defining the null model itself.
In some sense, a good but incomplete accordance can still indicate a quasi-stationary evolution, as long as the deviating patterns are always the same and systematically with the same amplitude. In this case, the network's dynamics are not entirely driven by that of the chosen constraints, as additional (in general unknown) explanatory properties would be required. 
By contrast, a network out of equilibrium (rigorously speaking, out of the quasi-equilibrium dynamics generated by the chosen constraints) would display wild and irregular deviations from the null model's expectations. 

\begin{figure}[t!]
\centering
\includegraphics[scale=0.6]{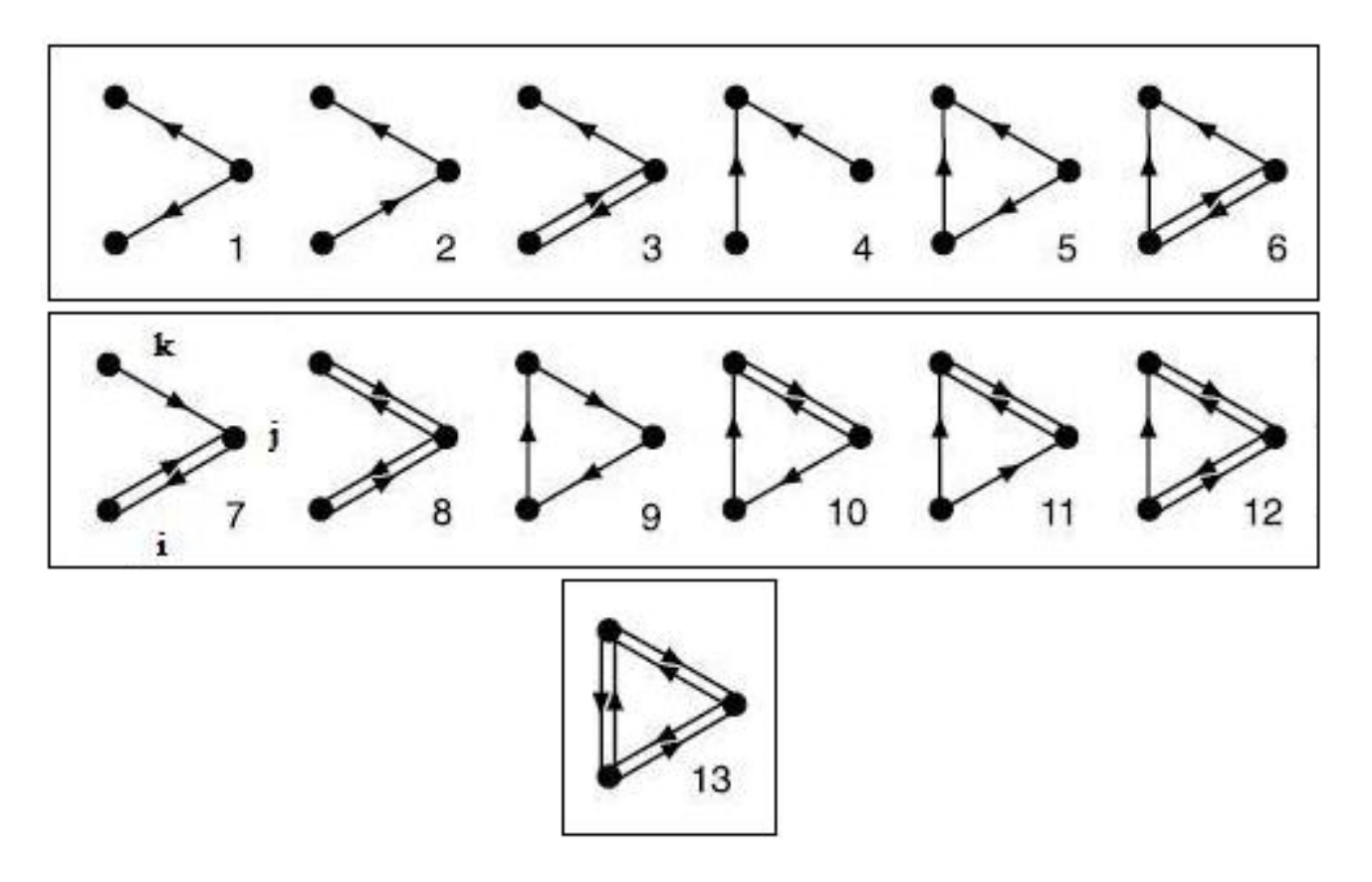}
\caption{The 13 triadic motifs representing all the possible non-isomorphic topological configurations involving three connected nodes in a directed network.}
\label{fig_mot}
\end{figure}

%In order to highlight deviations of the observed dynamics from the quasi-equilibrium ensemble, one should of course focus on higher-order properties not included in the constraints.
Since the constraints specified in the DCM and the RCM are node-specific %(e.g. the in- and out-degrees) 
and dyad-specific, %(e.g. the number of reciprocated or non-reciprocated dyads per vertex), 
the simplest non-trivial properties to monitor are \emph{triad-specific}, i.e. involving triples of nodes.
For this reason, in this paper we analyse in detail the so-called \emph{triadic motifs}  \cite{IEEEhowto:calda_book,IEEEhowto:foodwebmotifs,IEEEhowto:motifs,IEEEhowto:motifs2}, defined as the 13 non-isomorphic topological configurations involving three connected nodes in directed networks (see fig.\ref{fig_mot}).

The number $N_m$ of occurrences of a particular triadic motif $m$ (with $m=1\dots 13$) can be written in two equivalent ways. The first one employs products of adjacency matrix elements $\{a_{ij}\}$ and is suitable when using %the corresponding probabilities $\{p_{ij}\}$ of 
the DCM. 
The second one employs the  quantities $\{a^{\rightarrow}_{ij},\:a^{\leftarrow}_{ij},\:a^{\leftrightarrow}_{ij}\}$
and is particularly useful when using the %probabilities $\{p^{\rightarrow}_{ij},\:p^{\leftarrow}_{ij},\:p^{\leftrightarrow}_{ij}\}$ of 
the RCM. 
For example, the abundance of motif $m=10$ (see fig.\ref{fig_mot}) can be calculated as
\begin{eqnarray}
N_{10}(\mathbf{A})&=&\sum_{i\ne j\ne k}(1-a_{ij}) a_{ji} a_{ik} (1-a_{ki}) a_{jk} a_{kj}\nonumber\\
&=&\sum_{i\ne j\ne k}a_{ij}^{\leftarrow} a_{ik}^{\rightarrow} a_{jk}^{\leftrightarrow}
\label{eq:product}
\end{eqnarray}

\noindent and its expected values under the DCM and the RCM can be straightforwardly calculated as
\begin{equation}
\langle N_{10}\rangle_{DCM}=\sum_{i\ne j\ne k}(1-p_{ij}) p_{ji} p_{ik} (1-p_{ki}) p_{jk} p_{kj}
\label{eq:product2}
\end{equation}
and
\begin{eqnarray}
\langle N_{10}\rangle_{RCM}=\sum_{i\ne j\ne k}p_{ij}^{\leftarrow} p_{ik}^{\rightarrow} p_{jk}^{\leftrightarrow}
\label{eq:product3}
\end{eqnarray}
respectively. Given a real network $\mathbf{A}^*$, the usual way to compare the motifs' observed and expected abundance is defining the so-called \emph{z-scores}, i.e. the standardized quantities 
\begin{equation}
z_{m}\equiv\frac{N_{m}(\mathbf{A}^*)-\langle N_{m}\rangle^*}{\sigma^*[N_{m}]}
\label{eq:z}
\end{equation}
where $\sigma^*[N_{m}]\equiv\sqrt{\langle N_{m}^2\rangle^*-(\langle N_{m}\rangle^*)^2}$ is the standard deviation of $N_{m}$ under the null model.
If the observations were exactly reproduced, then the $z$-scores would be exactly zero. 
On the other hand, significantly large positive or negative $z$-scores indicate an over- or under-estimation of the motifs' empirical abundance respectively. 
For normally distributed variables, the deviations can be nicely quantified in terms of probability. In such cases, the intervals $z_m=\pm1,\:\pm2,\:\pm3$ select regions enclosing a probability of $68\%$, $95\%$ and $99.7\%$, respectively.
Choosing one of the above values as a threshold allows the identification of significantly deviating patterns.
For non-normally distributed variables, $z$-scores are still of value in highlight the most deviating patterns, even if they do not provide a clear probability estimation. 

When it is necessary to compare the $z$-scores of networks with different size, or of differently sized snapshots of the same network, one should bear in mind that the values of $z_m$ are sensitive to the number of nodes.
In order to consistently compare different snapshots, we need a size-independent measure. For this reason, it is customary to normalize the $z$-scores by introducing the \emph{significance profile} \cite{IEEEhowto:motifs,IEEEhowto:motifs2} defined as
\begin{equation}
SP_{m}\equiv\frac{z_{m}}{\sqrt{\sum_{m=1}^{13}z_{m}^2}}
\label{eq:SP}
\end{equation}

\noindent and measuring the \emph{relative} importance of each motif with respect to the other ones. Note that while the $z$-scores are unbounded quantities, the significance profile lies between $-1$ and $1$.

\section{Results and Discussion}
Equipped with the techniques and formalism described so far, we now show the results of the empirical analysis of the two economic networks mentioned in the Introduction%, i.e. the ITN and the DIN.
.

\subsection{The International Trade Network}
In the ITN, nodes represent world countries and a directed link from node $i$ to node $j$ represents the existence of an export relation from country $i$ to country $j$.
During the time period considered (1950-2000), the initial number (85) of countries roughly doubles, mainly because of many colonies becoming independent and the Soviet Uniot disgregating into many states.
Consequently, and also because of the globalization process, the number of links increases significantly \cite{IEEEhowto:mywtw}.
This implies that the local topological properties (degrees) of nodes vary considerably as well.
This circumstance makes the ITN an ideal example for testing whether an economic network undergoes a quasi-equilibrium evolution driven by the dynamics of the local properties.

\begin{figure}[t!]
\centering
\includegraphics[scale=0.625]{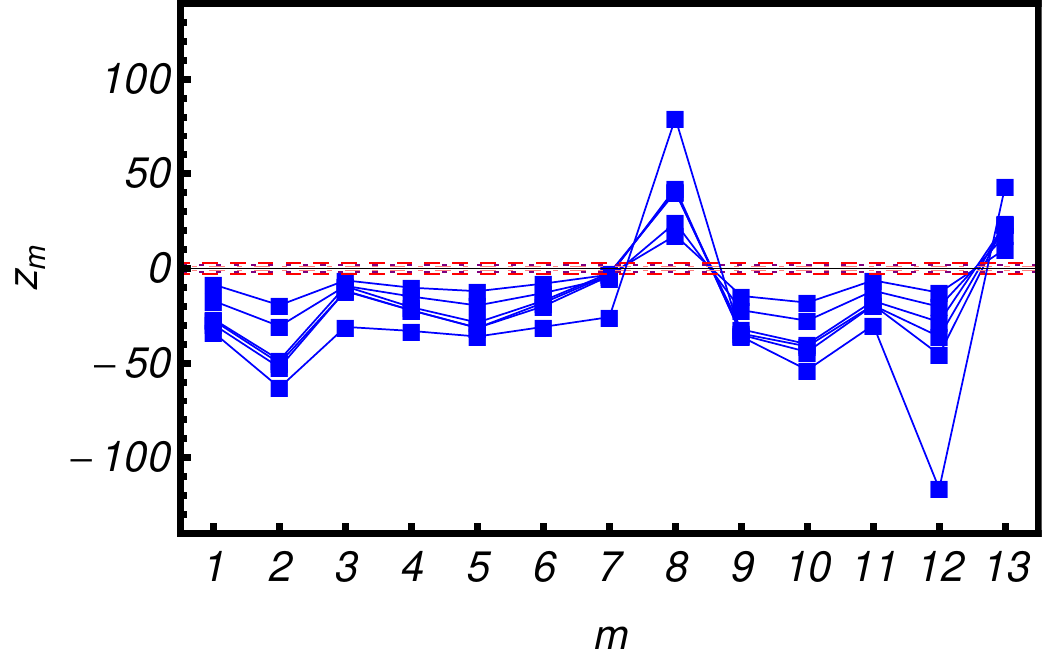}\\
\hspace{3.5mm}\includegraphics[scale=0.595]{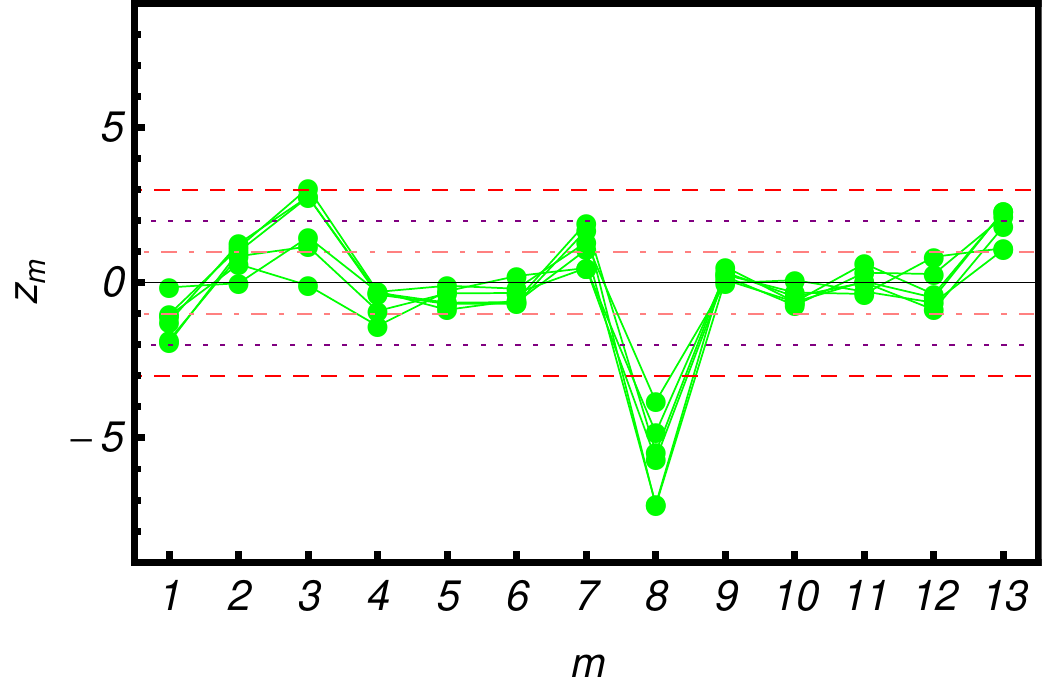}\\
\includegraphics[scale=0.62]{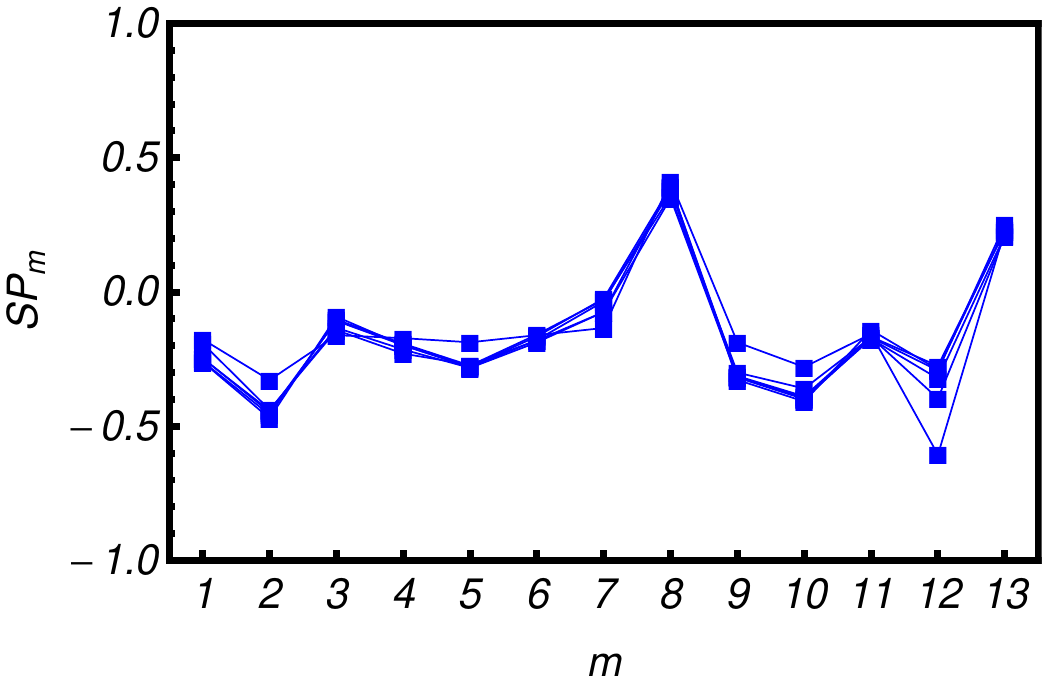}\\
\includegraphics[scale=0.62]{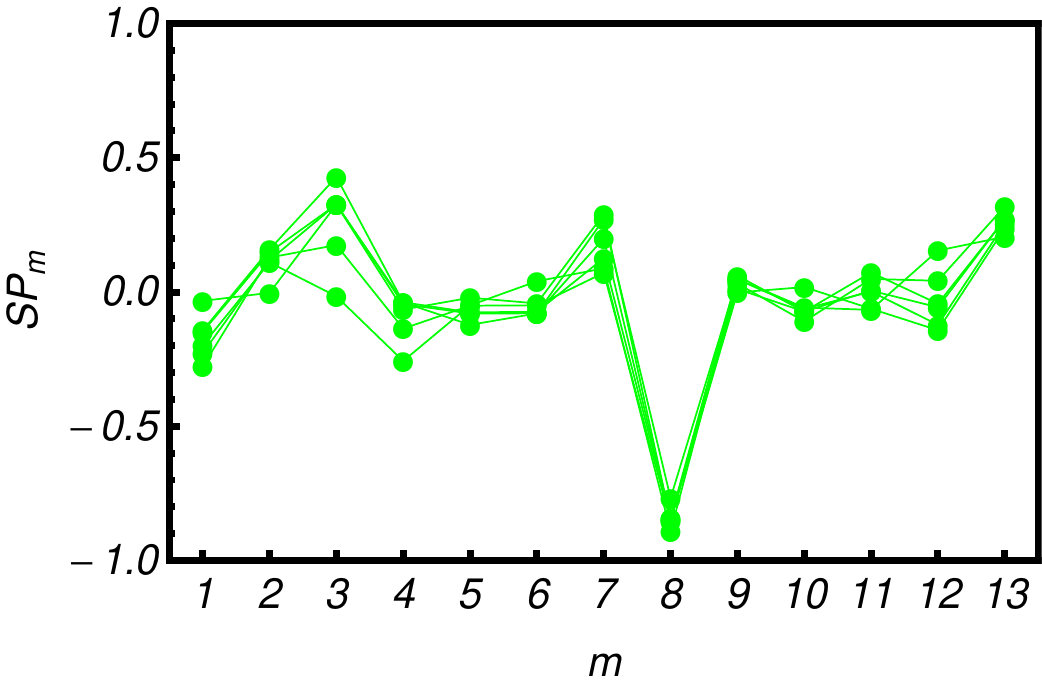}
\caption{$z$-scores (first and second panel) and significance profiles (third and fourth panel) of the 13 triadic, binary, directed motifs for the ITN in the years 1950, 1960, 1970, 1980, 1990 and 2000, under the DCM ($\textcolor{blue}{\blacksquare}$) and the RCM ($\textcolor{green}{\bullet}$). The dashed, red lines represent the values $z=\pm3$, the dotted, purple lines the values $z=\pm2$ and the dot-dashed, pink lines the values $z=\pm1$.}
\label{fig_wtw}
\end{figure}

%To check whether the changes in the structure of the ITN are almost completely explained by the changes in the degrees of nodes, we analyzed the $z$-scores of all the 13 triadic motifs illustrated in fig.\ref{fig_mot}, as defined in eq.(\ref{eq:z}), using both the DCM and the RCM.
The results of the anaysis of the $z$-scores, as defined in eq.(\ref{eq:z}), using both the DCM and the RCM. are shown in fig.\ref{fig_wtw}. 
We find that, under the DCM, the $z$-scores indicate large deviations between observations and expectations.
Moreover, the agreement worsens as the network evolves. 
These results are consistent with our analysis in ref. \cite{IEEEhowto:mymotifs}, and confirm that, while the some higher-order properties of the ITN were previously found to be well-reproduced by constraining the nodes' degrees \cite{IEEEhowto:pre1}, the triadic patterns are irreducible to (i.e. not explainable with) the in- and out-degrees themselves.
As we discuss in more detail below, it should in any case be noted that the shape of the profile of the $z$-scores displays a high degree of stability.

By contrast, under the RCM the agreement improves in a substantial way: now, all the $z$-scores (with the only exception of motif 8) lie within the error bars $z_m=\pm 3$. 
This indicates that, once reciprocated and non-reciprocated links are separately controlled for, the triadic structure of the network is almost completely explained.
Moreover, the shape of the profiles is more stable than under the DCM.
All these findings indicate that the reciprocity structure plays a strong role in shaping the topology of the ITN %, in line with previous results 
\cite{IEEEhowto:mywtw,IEEEhowto:mymotifs,IEEEhowto:myreciprocity}
%The information encoded in the reciprocity should therefore not be discarded when choosing the null model, 
and the RCM should be preferred to the DCM.

We now discuss in more detail the stationarity of the observed patterns.
Besides the $z$-scores, the panels of fig.\ref{fig_wtw} show the significance profiles for all 13 motifs, as defined in eq.(\ref{eq:SP}). 
We find that the global rescaling defining the significance profiles, by appropriately controlling for the changing size of the ITN, makes the curves for the 7 different snapshots collapse to a single trend.
This previously unreported (for the ITN) effect is obviously more evident under the DCM, since under the RCM the $z$-scores of the different snapshots were already largely overlapping. 
So, even if in absolute terms many structural quantities change (the number of nodes, the number of links, the degrees, etc.), we find that under both null models the significance profiles are extremely stable, i.e. the deviating patterns are systematic and the relative importance of each motif remains constant.

The above results indicate that the ITN is almost completely consistent with a quasi-equilibrium network driven by the local (non-)reciprocated degrees $k_{i}^{\rightarrow}$, $k_{i}^{\leftarrow}$ and $k_{i}^{\leftrightarrow}$.
The latter vary considerably over time, presumably mainly under the effect of complicated economic and political processes such as the creation of new independent states, the globalization and the establishment of reciprocated relationships.
However, once these processes are reabsorbed into the evolution of the local constraints, the quasi-equilibrium character of the network becomes manifest.

\subsection{The Dutch Interbank Network}
We now turn to the analysis of the DIN. In this network, nodes are Dutch banks and a link from node $i$ to node $j$ indicates that bank $i$ has an exposure larger than 1.5 M\euro, and with maturity shorter than one year, towards a creditor bank $j$. We consider 44 quarterly snapshots of the network, from the beginning of 1998 to the end of 2008.
The last year in the sample is when the recent financial and banking crisis became more manifest. 
During the evolution of the DIN, the number of banks and the number of connections (both total and per vertex) changed only moderately \cite{IEEEhowto:mybanks}.
The entity of the variation of these quantities in the DIN is therefore comparatively much smaller than in the evolution of the ITN described above.
This means that we might expect the DIN to display even more stable patterns than the ITN.
However, as we now show, the opposite is true.

\begin{figure}[t!]
\centering
\hspace{1mm}\includegraphics[scale=0.61]{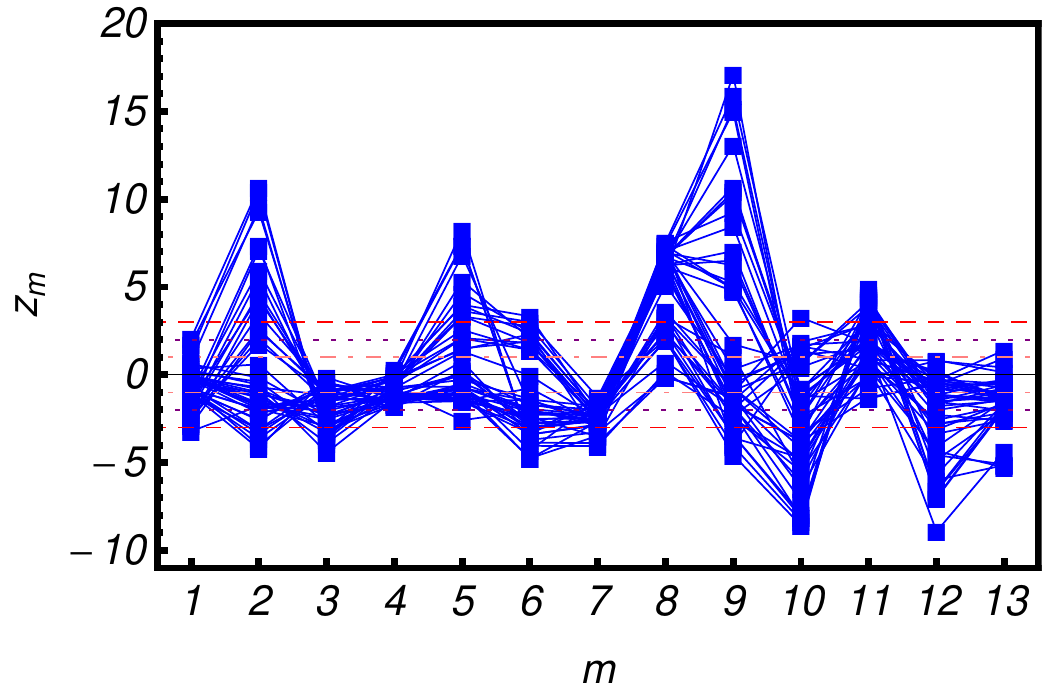}\\
\hspace{2mm}\includegraphics[scale=0.503]{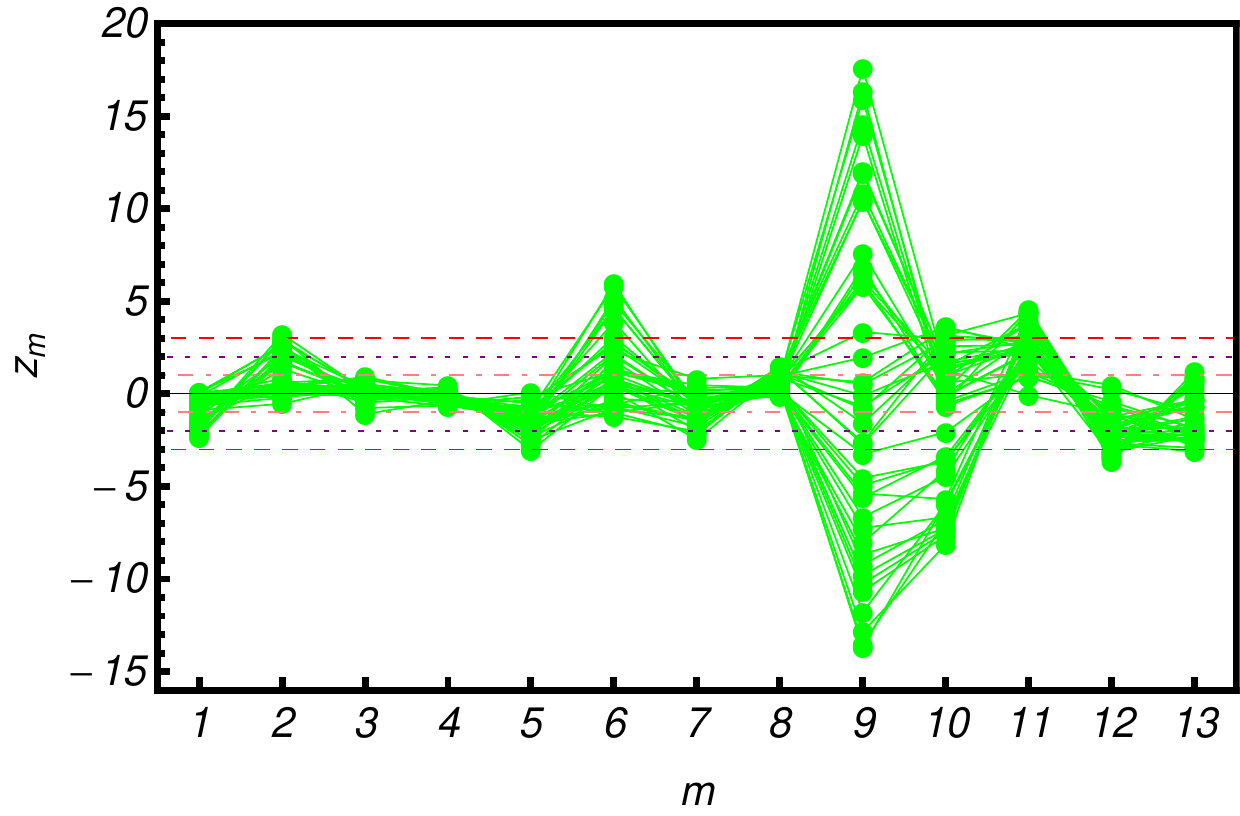}\\
\includegraphics[scale=0.62]{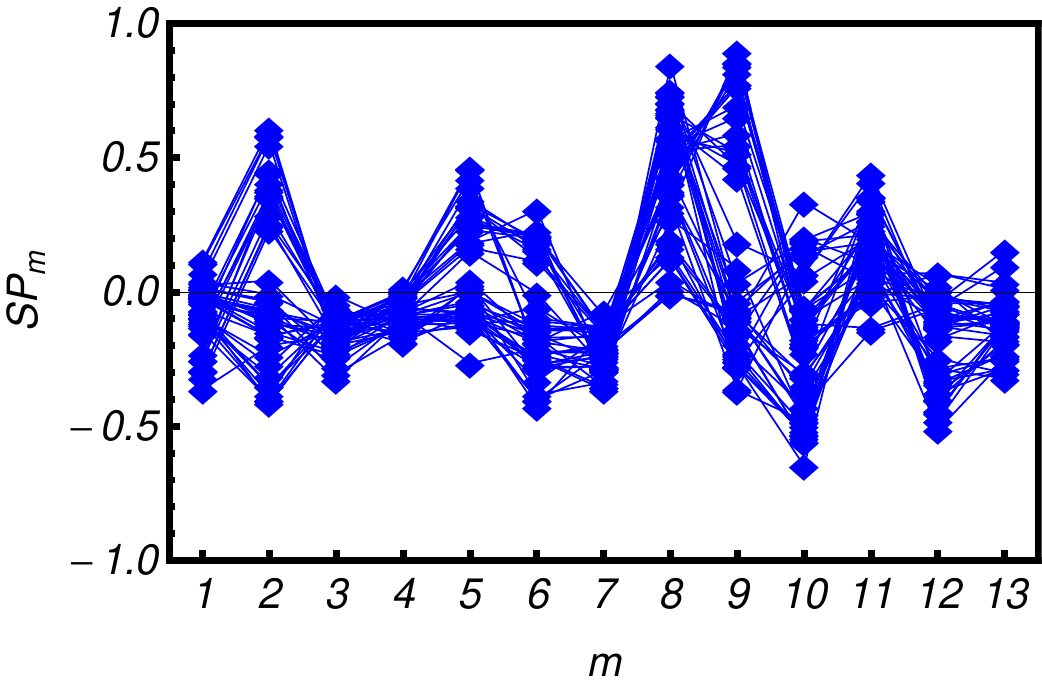}\\
\includegraphics[scale=0.62]{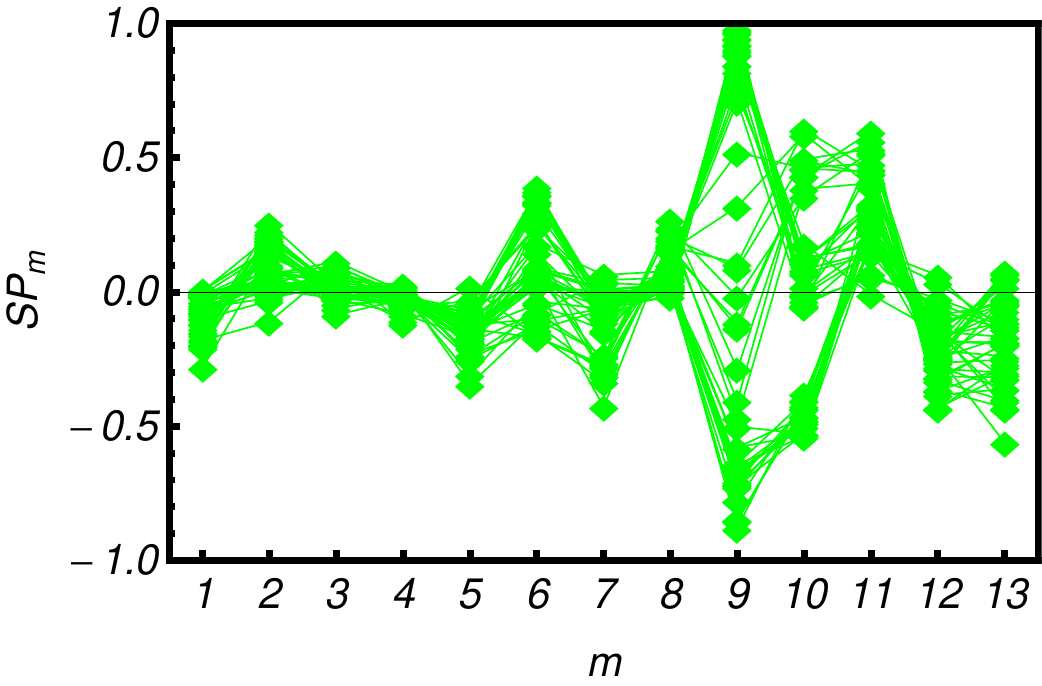}
\caption{$z$-scores (first and second panel) and significance profiles (third and fourth panel) of the 13 triadic, binary, directed motifs for the 44 quarterly snapshots of the DIN between 1998 and 2008, under the DCM ($\textcolor{blue}{\blacksquare}$ and $\textcolor{blue}{\blacklozenge}$) and the RCM ($\textcolor{green}{\bullet}$ and $\textcolor{green}{\blacklozenge}$). The dashed, red lines represent the values $z=\pm3$, the dotted, purple lines the values $z=\pm2$ and the dot-dashed, pink lines the values $z=\pm1$.}
\label{fig_din1}
\end{figure}

If we repeat the calculation of the $z$-scores and significance profiles that we used to produce fig.\ref{fig_wtw}, for the DIN we obtain the corresponding fig.\ref{fig_din1}. 
What we find is that, unlike the ITN, the DIN displays highly non-stationary profiles, with different snapshots never collapsing to a unique curve.

Many motifs have, in different periods, both positive and negative $z$-scores, indicating a complete inversion of their significance (from under-representation to over-representation and vice versa). 
Note that, since the number of nodes remains approximately constant \cite{IEEEhowto:mybanks},
in this case the significance profiles do not appreciably change the shape of the $z$-scores. 
This confirms that the evolution of the triadic profiles is not due to changes in the size of the network, and is a genuine effect.

\begin{figure}[t!]
\centering
\includegraphics[scale=0.62]{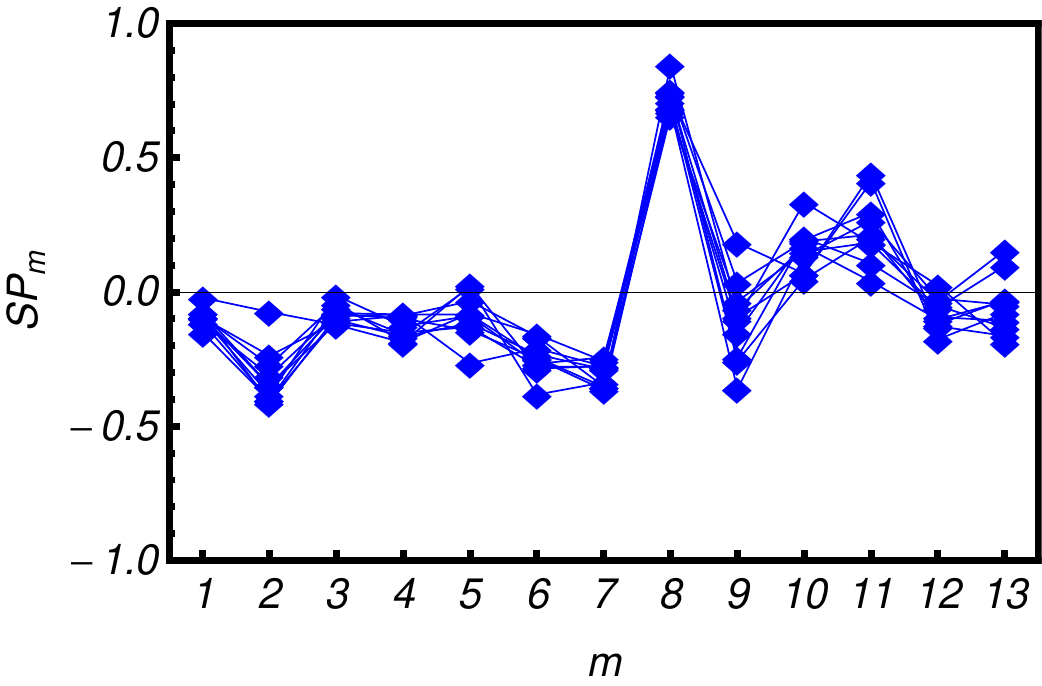}\\
\includegraphics[scale=0.62]{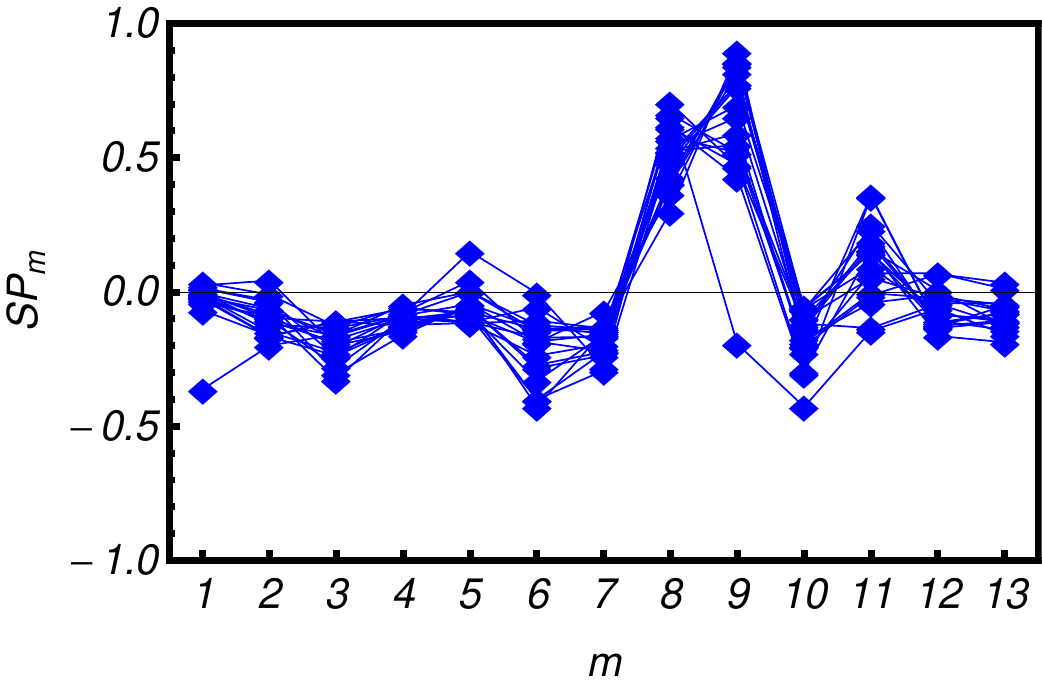}\\
\includegraphics[scale=0.62]{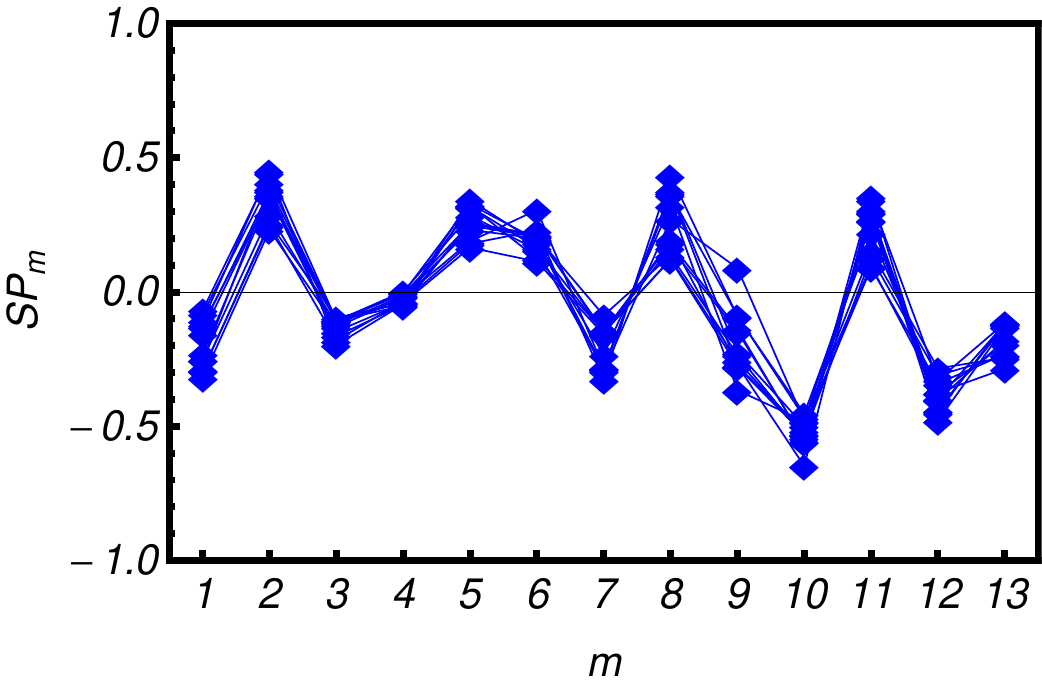}\\
\includegraphics[scale=0.62]{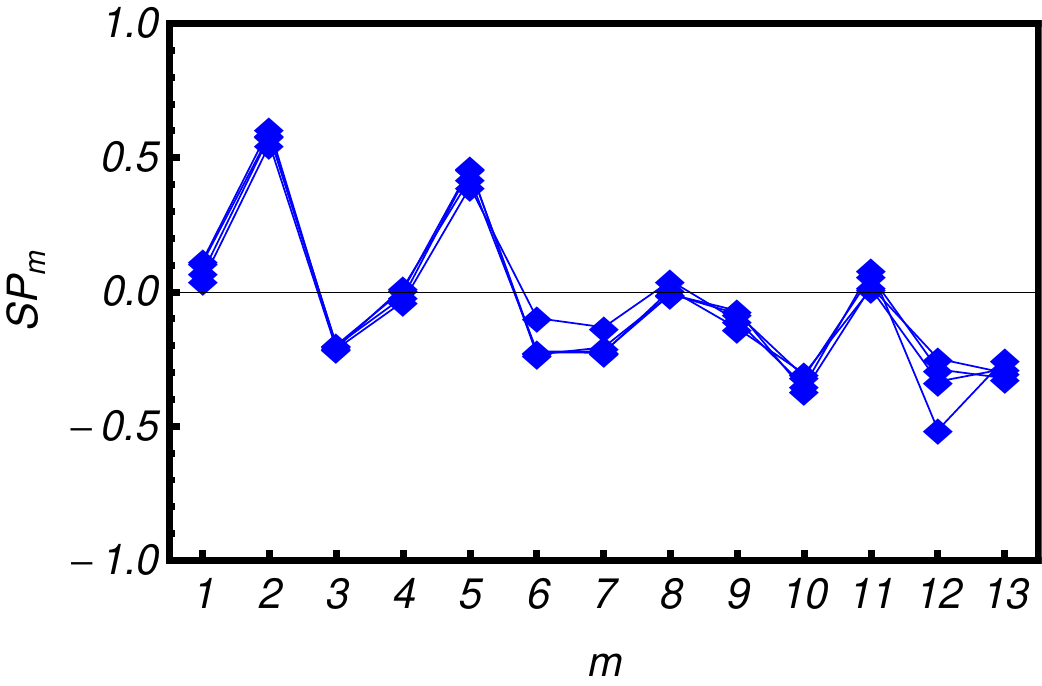}
\caption{Significance profiles of the DIN under the DCM ($\textcolor{blue}{\blacklozenge}$) for the four subperiods (from top to bottom) 1998Q1-2000Q2, 2000Q3-2004Q4, 2005Q1-2007Q4 and 2008Q1-2008Q4.}
\label{fig_din2}
\end{figure}

\begin{figure}[t!]
\centering
\includegraphics[scale=0.62]{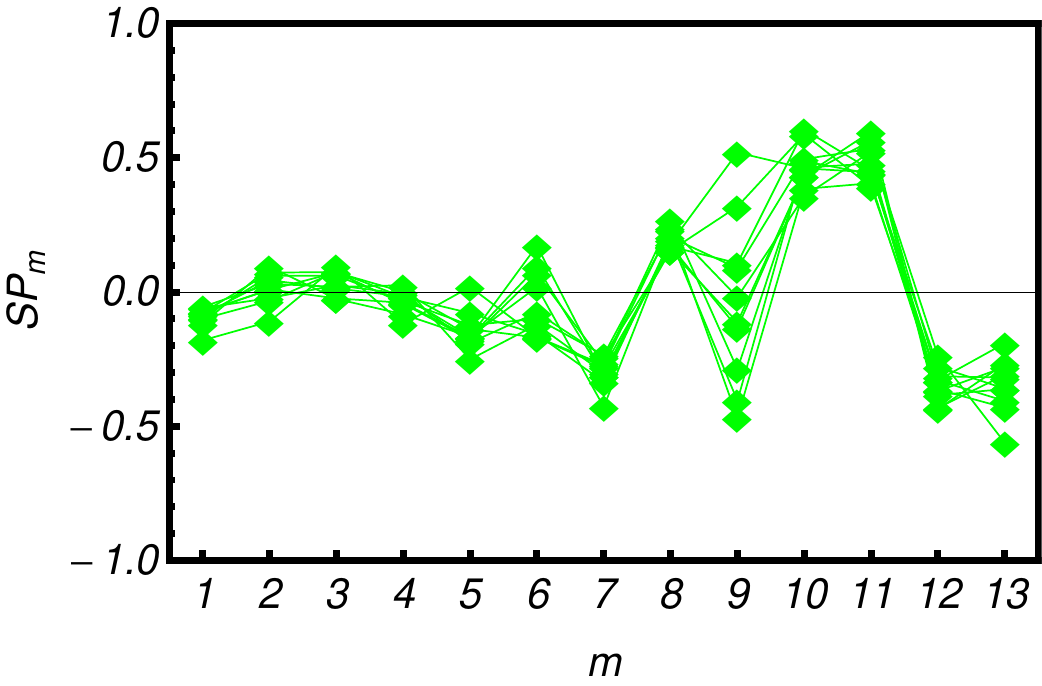}\\
\includegraphics[scale=0.62]{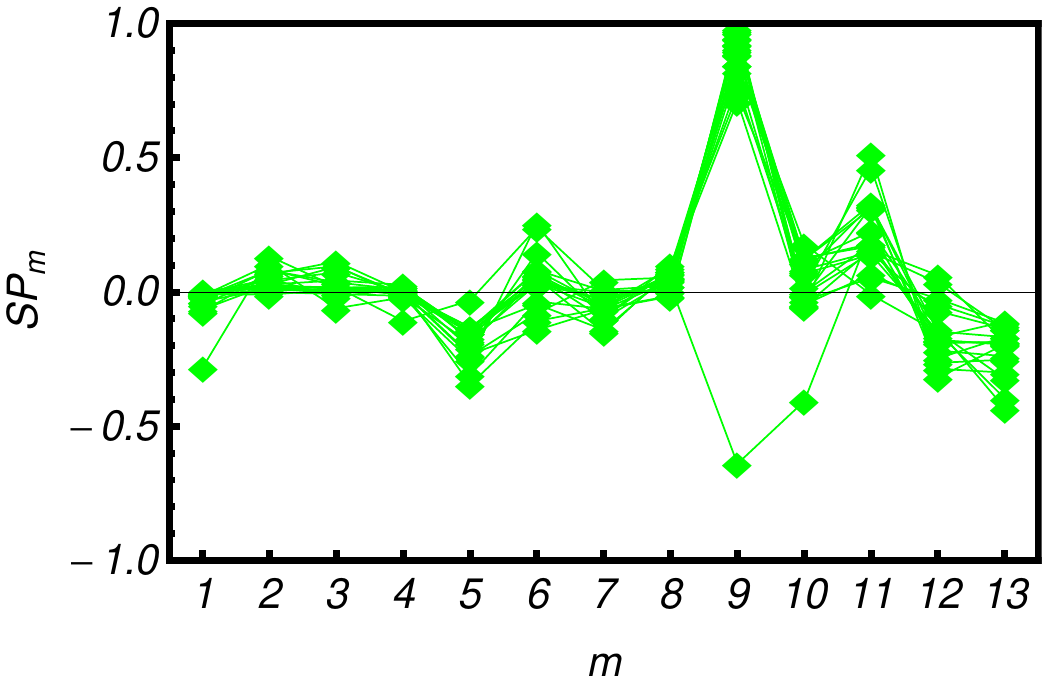}\\
\includegraphics[scale=0.62]{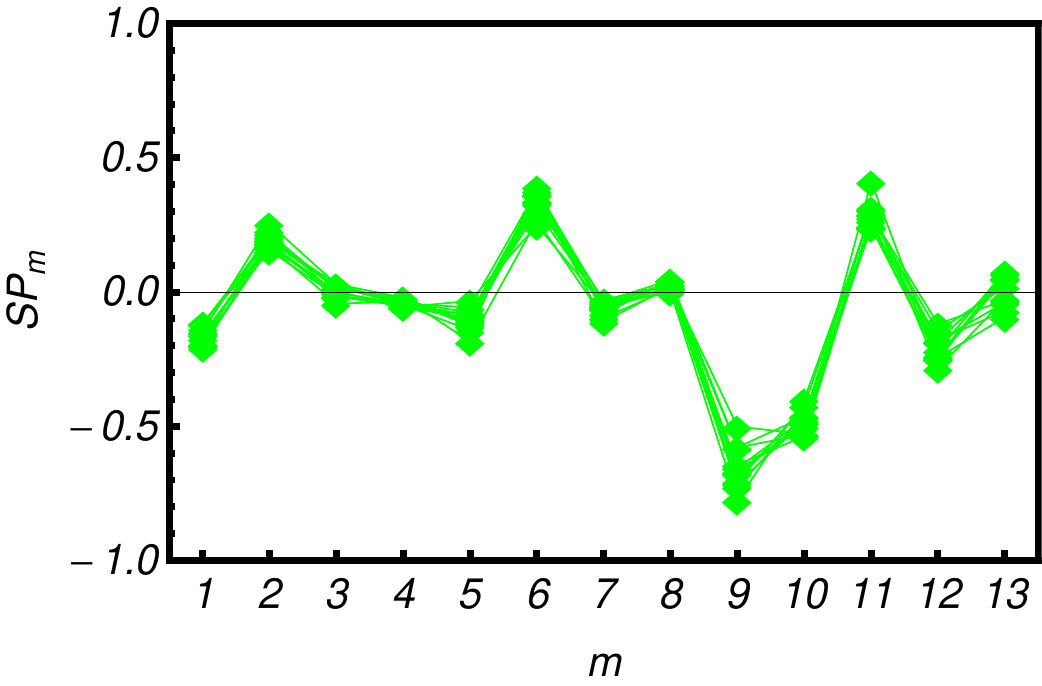}\\
\includegraphics[scale=0.62]{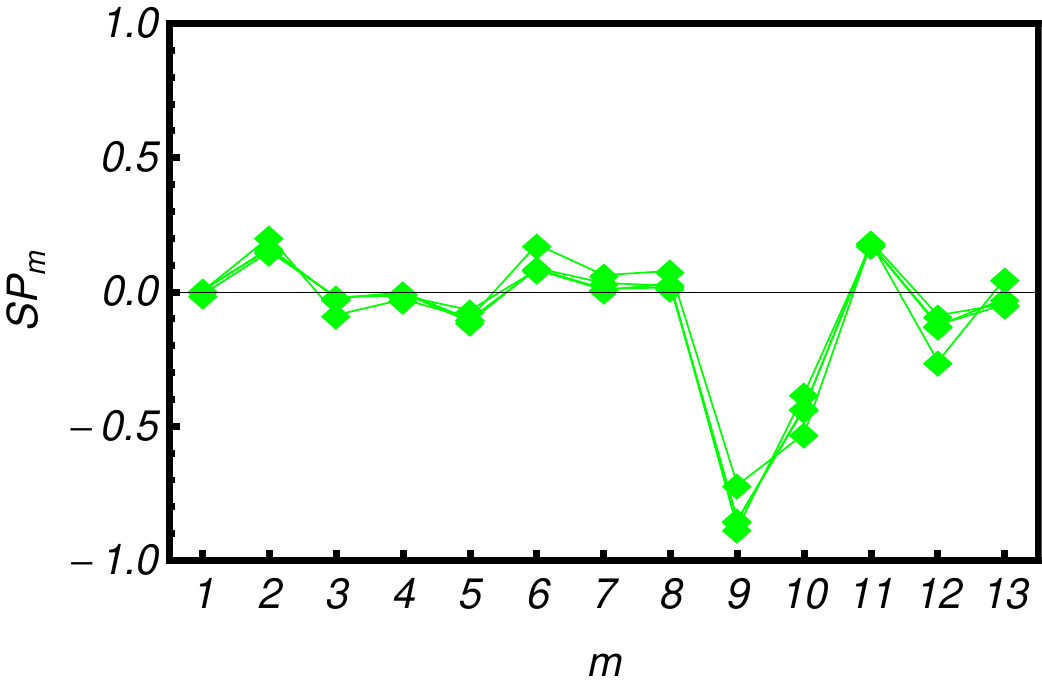}
\caption{Significance profiles of the DIN under the RCM ($\textcolor{green}{\blacklozenge}$) for the four subperiods (from top to bottom) 1998Q1-2000Q2, 2000Q3-2004Q4, 2005Q1-2007Q4 and 2008Q1-2008Q4.}
\label{fig_din3}
\end{figure}

The large (in absolute value) $z$-scores and their wild temporal fluctuations indicate that, unlike the ITN, the DIN behaves like an out-of-equilibrium network, whose driving dynamics cannot be captured by the selected constraints alone.
However, what is most interesting is the presence, under both null models, of four different shapes of the triadic profiles, subdividing the analysed decade into four subperiods (1998Q1-2000Q2, 2000Q3-2004Q4, 2005Q1-2007Q4, 2008Q1-2008Q4, where $x$Q$i$ denotes the $i$th quarter of  year $x$).
Inside each of these four subperiods, the significance profiles are largely stable.
This is shown in fig.\ref{fig_din2} under the DCM and in fig.\ref{fig_din3} under the RCM. 
Both figures show the four subperiods separately, and the almost complete collapse of all snapshots within each subperiod. 

The above results indicate that the overall non-stationary dynamics of the DIN can be approximately decomposed into four relatively stationary phases connected by major structural transitions.
Importantly, the fourth subperiod exactly coincides with the year 2008, i.e. the year when the global interbank crisis became manifest. 
So the triadic profiles for this particular period portray in some sense the `topological fingerprints' of the crisis.
It is therefore interesting to notice that these fingerprints were to a large extent anticipated by the significance profiles of the preceding subperiod (the third one) starting in 2005.
This confirms our recent finding that the some dyadic and triadic properties characterizing the DIN during the 2008 crisis, if identified using a uniform null model controlling only for the overall network density, appeared to suddenly collapse to their final values only when the crisis was already manifest; by contrast, if identified using the DCM or the RCM, the same quantities showed a gradual evolution towards the collapsed configuration \cite{IEEEhowto:mybanks}. 

The third subperiod (2005-2007) therefore coincides with a latent `pre-crisis' phase. It is indeed remarkable that the most dramatic change of the significance profiles occurs precisely between the second and third subperiods, and not between the third and fourth ones (as naively expected). 
This is an additional indication that the main structural transition occurred at the beginning of the pre-crisis phase, and not at the onset of the crisis itself.
This suggests that monitoring the non-stationary evolution of the triadic profiles could potentially represent a way to detect early-warning signal of interbank crises. 
These considerations appear to indicate that the non-stationary character of the DIN on one hand makes the description of the system more complicated than that of equilibrium networks such as the ITN, but on the other hand might be precisely what is needed in order to detect early-warning signals.

\section{Conclusions}

In the present paper we have proposed a way to investigate whether real economic network are in or out of equilibrium. 
We introduced the concept of quasi-equilibrium graph ensembles driven by the dynamics of local constraints.
This allowed us to relate the stationarity of a network to its \emph{statistical typicality} with respect to a chosen ensemble. 
So, as evident from the results of the analysis, `stationary' here does not mean that the numerical values of certain topological quantities stay constant across time: it means that the newtork's evolution is systematically driven by the dynamics of the chosen constraints, and so by the (presumably relatively simple) process determining the evolution of the constraints themselves. 
So, even if the main quantities usually investigated in network theory (the number of nodes, the number of connections, the nodes' degree, etc.) vary over time, the explanatory power of the chosen constraints may remain constant.
Our empirical results show that the two systems considered in our analysis display two completely different behaviours: while the ITN is an equilibrium network, the DIN is an out-of-equilibrium one.

In the ITN, while the number of incoming and outgoing connections of all vertices encloses the necessary information to reproduce the properties like the degree-degree correlations and the clustering coefficient \cite{IEEEhowto:pre1}, it always fails in explaining the triadic structure. 
On the other hand, we found that the RCM, which also constrains the numbers of reciprocated links, can replicate the triadic structure almost perfectly.% This confirms the important role of reciprocity in economic networks.

In the DIN (under both null models) it turns out that the 44 temporal snapshots do not collapse to a single profile, and that four subperiods with different profiles can be distinguished. 
The major topological changes at the beginning of 2005 appear to mark the start of an early-warning pre-crisis phase, eventually evolving towards the interbank crisis of 2008.
These results call for future studies aimed at understanding the potential of monitoring the non-stationary  properties of interbank networks within the framework of bank regulation.

\section*{Acknowledgments}

D. G. acknowledges support from the Dutch Econophysics Foundation (Stichting Econophysics, Leiden, the Netherlands) with funds from beneficiaries of Duyfken Trading Knowledge BV, Amsterdam, the Netherlands.



%\bibliographystyle{IEEEtran}

\end{document}